\documentclass[
    ,final            
  ]
  {aipproc}

\layoutstyle{6x9}
\usepackage{graphicx}
\usepackage{psfig}
\usepackage{amsmath}

\begin{document}

\title{The iron $K_\alpha$-line as a tool for an evaluation of black hole parameters}

\author{A.F. Zakharov}
{address={Institute of Theoretical and Experimental Physics, Moscow \& \\
Astro Space Center of Lebedev Physics Institute, Moscow, Russia}}



\begin{abstract}
  Recent X-ray observations of microquasars and Seyfert galaxies
reveal  broad emission lines in their spectra, which can arise in
the innermost parts of accretion disks. Simulations indicate that
at low inclination angle the line is measured by a distant
observer as characteristic two-peak profile. However, at high
inclination angles ($> 85^0$) two additional peaks arise. This
phenomenon was discovered by Matt et al. (1993) using the
Schwarzschild black hole metric to analyze  such effect. They
assumed that the effect is applicable to a Kerr metric far beyond
the range of parameters that they exploited. We check and confirm
their hypothesis about such a structure of the spectral line shape
for the Kerr metric case. We use no astrophysical assumptions
about the physical structure of the emission region except the
assumption that the region should be narrow enough.
 Positions and heights of these extra peaks drastically
depend on both the radial coordinate of the emitting region
(annuli) and the inclination angle.   It was found that these
extra peaks arise due to gravitational lens effect in the strong
gravitational field, namely they are formed by photons with some
number of revolutions around black hole. This conclusion is based
only on relativistic calculations without any assumption about
physical parameters of the accretion disc like X-ray surface
emissivity etc.
 We discuss how analysis of the iron spectral line
shapes could give an information about an upper limit of magnetic
field near black hole horizon.

\end{abstract}

\maketitle

\section{SIGNATURES OF HIGHLY INCLINATED ACCRETION DISKS NEAR GBHCs AND AGNs}


More than ten years ago it was predicted that profiles of lines
emitted by AGNs and X-ray binary systems\footnote{Some of them are
microquasars (for details see, for example, \cite{Grein99, Mira00,
Mira02a}).} could have an asymmetric double-peaked shape (e.g.
\cite{Chen89,Fabian89,MPS93}).
Generation of the  broad $K_\alpha$ fluorescence lines as a result
of irradiation of a cold accretion disk was discussed by many
authors (see, for example,
\cite{MPPS92,MPPS92a,Matt92,MFR93,Bao93,Mart02} and references
therein).
\citet{Popov01,Popov02} discussed influence of microlensing on the
distortion of spectral lines including Fe $K_\alpha$ line, that
can be significant in some cases. \citet{ZPJ03} showed that the
optical depth for microlensing could be significant for
cosmological distributions of microlenses.
    Recent X-ray observations of Seyfert galaxies, microquasars
and binary systems
(\cite{fabian1,tanaka1,nandra1,nandra2,malizia,sambruna,
yaqoob4,ogle1,Miller02} and references therein) confirm these
considerations in general and reveal broad emission lines in their
spectra with characteristic two-peak profiles. A comprehensive
review by \citet{Fabi00} summarizes the detailed discussion of
theoretical aspects of possible scenarios for generation of broad
iron lines in AGNs. These lines are assumed to arise in the
innermost parts of the accretion disk, where the effects of
General Relativity (GR) must be taken into account, otherwise it
appears very difficult to find a natural explanation for observed
line profile.

     Numerical simulations of the line structure
are be found in a number of papers
\cite{Bao93,Koji91,Laor91,Bao92,BHO94,bromley,Fan97,pariev2,pariev1,pariev3,
Rusz00,Ma02}. They indicate that the accretion disks in Seyfert
galaxies are usually observed at the inclination angle $\theta$
close to $30^0$ or less. This occurs because according to the
Seyfert galaxy models, an opaque dusty torque surrounds the
accretion disk which does not allow us to observe the disk at
larger inclination angles.

     However, at inclination angles $\theta > 80^0$, new
observational manifestations of GR could arise. (\citet{MPS93}
discovered such phenomenon for a Schwar\-z\-s\-child black hole,
moreover the authors predicted that their results could be
applicable  to a Kerr black hole over the range of parameters
exploited). The authors mentioned that this problem was not
analyzed in detail for a Kerr metric case and it would be
necessary to investigate this case. Below we do not use a specific
model on surface emissivity of accretion (we only assume that the
emitting region is narrow enough). But general statements (which
will be described below) can be generalized to a wide disk case
without any problem. Therefore, in this paper we check and confirm
their hypothesis for the Kerr metric case and for a
Schwar\-z\-s\-child black hole using other assumptions about
surface emissivity of accretion disks. In principle, such a
phenomenon could be observed in microquasars and X-ray binary
systems where there are neutron stars and black holes with stellar
masses.

\subsection{Numerical methods}

We used an approach  discussed in detail in papers
\cite{zakh91,zakharov01,
zakharov1,zak_rep1,zak_rep2,Zak02pr,zak_rept,Zak_rep02_xeus,
Zak_rep02_Gamma,Zak_Rep03_Lom,Zakharov_Repin_Ch03,Zak03_Sak,Zakharov_Repin_Pom04,Zak_Rep03a,Zak_Rep03b,zak_rep03,ZKLR02}.
The approach was used in particular to simulate   spectral line
shapes. For example, \citet{ZKLR02} used this approach to simulate
the influence of a magnetic field on spectral line profiles. This
approach is based on results of a qualitative analysis (which was
done by  for different types of geodesics near a Kerr black hole
\cite{zakh86,zakh89}).
 The equations of photon motion in the Kerr metric are reduced to
the following system of ordinary differential equations in
dimensionless Boyer -- Lindquist coordinates
\cite{zakh91,zakharov1}:
\begin{eqnarray}
   \frac{dt}{d\sigma}
                           & = &
      - a \left(a \sin^2\theta - \xi\right) +
      \frac{r^2 + a^2}{\Delta}
       \left(r^2 + a^2 - \xi a\right),
                        \label{eq1}                       \\
   \frac{dr}{d\sigma} & = & r_1,       \label{eq2}        \\
   \frac{dr_1}{d\sigma}    & = &
      2r^3 + \left(a^2 - \xi^2 - \eta\right) r +
      \left(a - \xi\right)^2 + \eta,                      \\
   \frac{d\theta}{d\sigma} & = & \theta_1,                \\
   \frac{d\theta_1}{d\sigma}
                           & = &
      \cos\theta \left(\frac{\xi^2}{\sin^3\theta} -
                       a^2 \sin\theta
                 \right),              \label{eq5}       \\
   \frac{d\phi}{d\sigma}   & = &
      - \left(a - \frac{\xi}{\sin^2\theta}\right) +
      \frac{a}{\Delta}
           \left(r^2 + a^2 - \xi a \right),
                     \label{eq6}
\end{eqnarray}
where $\Delta=r^2-2r+a^2$; $\eta = Q/M^2E^2$ and $\xi = L_z/ME$
are the Chan\-dra\-sekhar constants \cite{chandra} which are
derived from the initial conditions of the emitted quantum in the
disk plane.
 The system~(\ref{eq1})-(\ref{eq6}) has two first
integrals
\begin{eqnarray}
  \epsilon_1 & \equiv & r_1^2 - r^4 -
      \left(a^2 - \xi^2 - \eta\right) r^2 -    
      2\left[\left(a - \xi\right)^2 + \eta \right] r +
      a^2\eta = 0,                  \\
  \epsilon_2 & \equiv & \theta_1^2 - \eta - \cos^2\theta
      \left(a^2 - \frac{\xi^2}{\sin^2\theta}\right) = 0,
                    \label{eq8}
\end{eqnarray}
which can be used for the accuracy control of computation.

     Solving Eqs. (\ref{eq1})--(\ref{eq6}) for
monochromatic quanta emitted by a ring we can calculate a spectral
line shape $I_\nu(r,\theta)$ which is registered by a distant
observer at inclination angle $\theta$.

\subsection{Disk  model}

    To simulate the structure of the emitted line it is necessary
first to choose a model for the emissivity of an accretion disk.
We exploit two different models, namely  we consider a narrow and
thin disk moving in the equatorial plane near a Kerr black hole as
the first model and as  we analyze the inner wide part of an
accretion disk with a temperature distribution which is chosen
according to the \cite{shasun,LipShak} with fixed inner and outer
radii $r_i$ and $r_o$ as the second model.
 Usually a power  law is
used  for  wide disk emissivity (see, for example,
\cite{Laor91,MM96,MKM00,MMK02}). However, other models for
emissivity can not  be excluded for such a wide class of accreting
black holes, therefore,  to demonstrate how another emissivity law
could change line profiles we investigate  such a emissivity law.

First, we assume that the source of the emitting quanta is a
narrow thin disk rotating in the equatorial plane of a Kerr black
hole. We also assume that the disk is opaque to radiation, so that
a distant observer situated on one disk side cannot measure the
quanta emitted from its other side.

For second case of the Shakura -- Sunyaev disk model, we assume
that the local emissivity is proportional to the surface element
and and $T^4$, where $T$ is a local temperature.
     The emission intensity of the ring is proportional to its area.
The area of the emitting ring (width $dr$) differs in the Kerr
metric from its classical expression $dS = 2\pi rdr$  and should
be replaced with
\begin{equation}
   dS = \frac{2\pi \left(r^2 + a^2\right)}
             {\displaystyle\sqrt{r^2 - rr_g + a^2\phantom{1}}}\, dr.
          \label{eq9}
\end{equation}

     For simulation purposes we assume that
the emitting region lies entirely in the innermost region of the
$\alpha$-disk (zone~$a$) from $r_{out} = 10\,r_g$ to $r_{in} =
3\,r_g$ and the emission is monochromatic in the co-moving
frame.\footnote{We use as usual the notation $r_g=2GM/c^2$.}
 The frequency of this emission set as  unity
by convention.

\subsection{Simulation results}

\begin{figure*}
\vspace{-.5cm}
 \psfig{figure=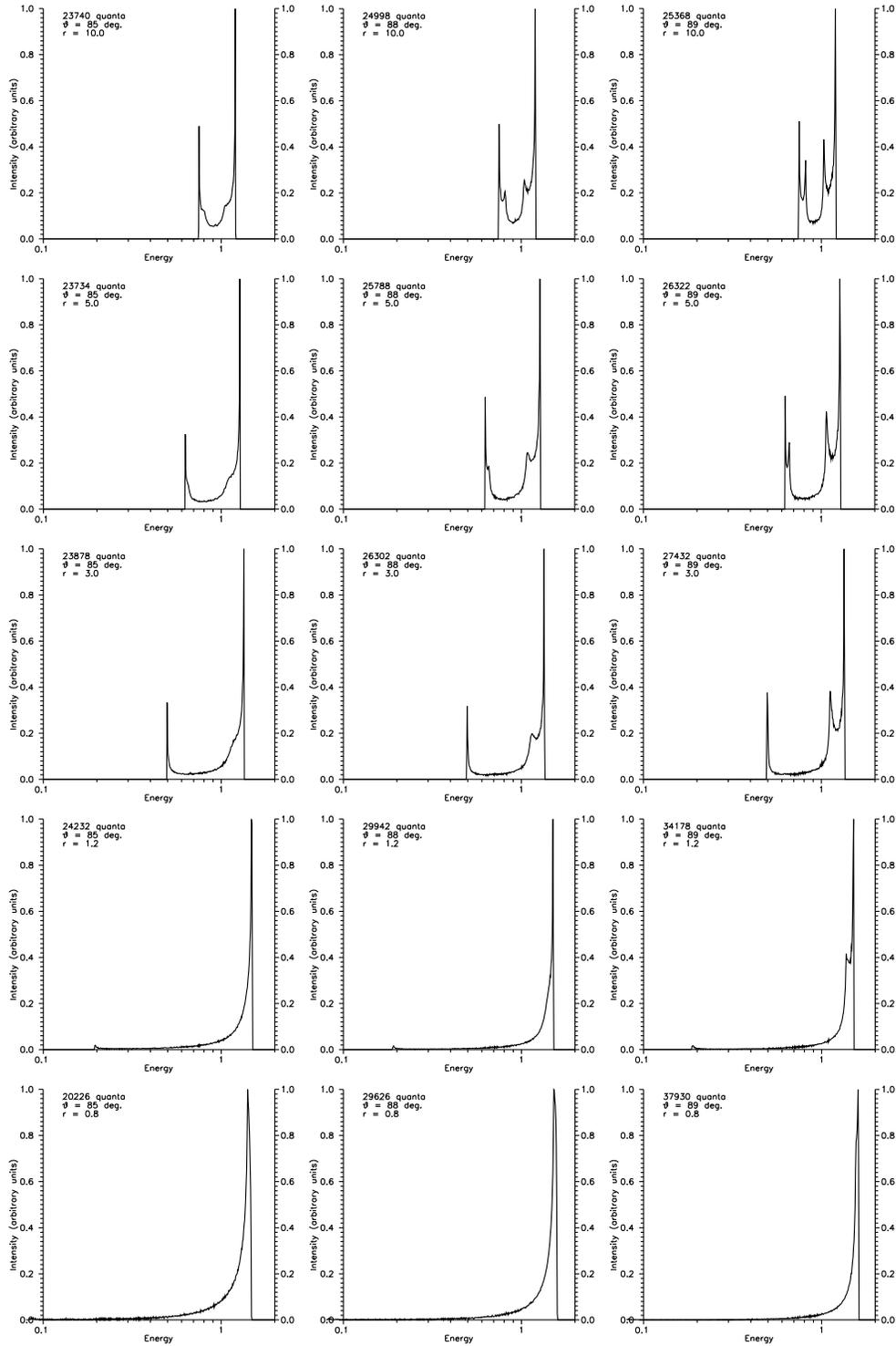,width=0.88\textwidth}
\vspace{-1cm}
  \caption{Line profiles  for
           high inclination angles, $\theta > 85^0$, due to
           gravitational lens effects in the strong gravitational
           field approach. The Kerr (rotation) parameter was
           chosen as $a = 0.9981$. The radii decrease
           from top to bottom, the inclination angles increase from
           left to right; their values are shown in each panel along
           with the number of quanta included in the spectrum.}
  \label{peaks01}
\end{figure*}

      Spectral line profiles of a narrow ring observed at large inclination
angles $\theta > 85^0$ and different radii $r$ are shown in
Fig.~\ref{peaks01} (a Kerr metric generalization of Fig. 1 from
\cite{MPS93} which was drawn using calculations for the
Schwarzschild case). The ring is assumed to move in the equatorial
plane of a Kerr black hole with an almost extreme rotation
parameter $a = 0.9981$. The inclination angle increases  from left
to right and the radial coordinate from bottom to top. The figure
indicates that there are  practically no new specific features of
profiles, thus,  the line profile remains one-peaked with a
maximum close to $1.6\,E_{lab}$ and a very long red wing without
any significant details.

For the lowest radii there are no signatures of multiple peaks of
spectral line shapes even for high inclination angles (the bottom
row in Fig. 1 which corresponds to $r=0.8\,r_g$).

  Increasing the radius to $r = 1.2\, r_g$, an  additional
blue peak arises in the vicinity of the blue maximum at the
highest inclination angle $\theta = 89^0$. The red maximum is so
small that no details can be distinguished in its structure. At
lower inclination angles $\theta \le 88^0$ the blue maximum also
has no details and the entire line profile remains essentially
one-peaked.

    For $r = 3\, r_g$  additional details in the blue peak
appear for $\theta \ge 85^0$. Thus, for $\theta = 85^0$ we have a
fairly clear bump, at $\theta = 88^0$ it changes  into a small
complementary maximum and for $\theta = 89^0$ this maximum becomes
well-distinguished. Its position in the last case differs
significantly from the main maximum: $E_3 = 1.12\,E_{lab}$, $E_4 =
1.34\,E_{lab}$.

    When further increasing the radius the red maximum also
bifurcates. This effect becomes  visible  for $r = 5\,r_g$ and
$\theta \ge 88^0$.  Thus, for $\theta = 85^0$ we have only a
faintly discernible (feebly marked) bump, but for $\theta = 88^0$
both complementary maxima (red and blue) arise. For $\theta =
89^0$ we have already four maxima in the line profile: $E_1 =
0.63\,E_{lab}$, $E_2 = 0.66\,E_{lab}$, $E_3 = 1.07\,E_{lab}$, $E_4
= 1.28\,E_{lab}$. Note that the splitting for the blue and red
maxima is not equal, moreover, $E_4 - E_3 \approx 7\, \left(E_2 -
E_1\right)$.

    For $r = 10\, r_g$ the profile becomes more narrow,
but the complementary peaks appear very distinctive. We have a
four-peak structure for  $\theta \ge 88^0$. It is interesting to
note that for $\theta = 89^0$ the energy of the blue complementary
peak is close to its laboratory value.

   Note that the effect almost disappears when the radial coordinate
becomes less than $r_g$, i.e. for the orbits which could exist
only near a Kerr black hole.

\begin{figure*}
  \psfig{figure=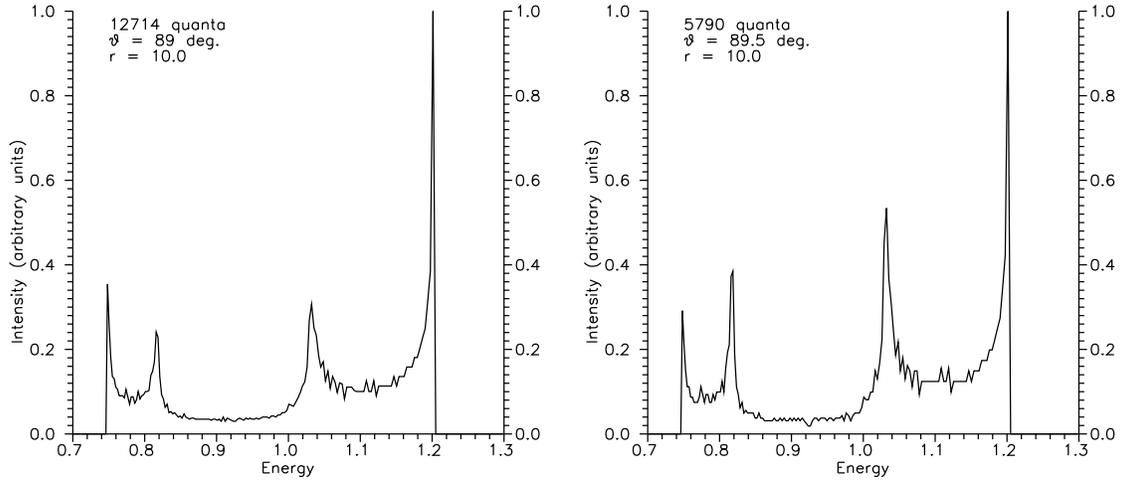,width=\textwidth}
  \caption{Details of a hot spot line profile for the most
           distinctive case with $r=10\,r_g$ and $a = 0.9981$
           (see the top row of Fig.~\ref{peaks01}).
           The images of all orders are counted. The left panel includes
           all the quanta, registered at infinity with $\theta > 89^0$.
           The right panel includes the quanta with $\theta > 89.5^0$.}
  \label{peaks02}
\end{figure*}

\begin{figure}
  \includegraphics[width=0.8\textwidth]{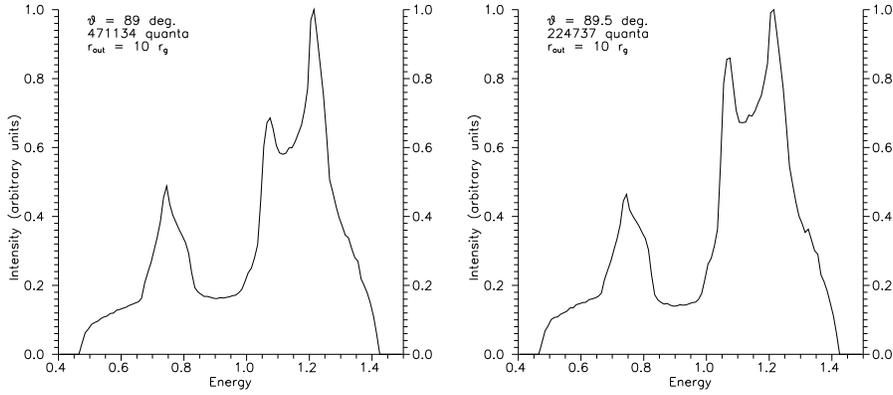}
  \caption{Details of the line structure for an $\alpha$-disk in
           the Schwarzschild metric with outer and inner edges of
           emitting region equal to $r_{out} = 10\,r_g$ and
           $r_{in} = 3\,r_g$, respectively. The left panel includes all
           the quanta, registered at infinity with $\theta > 89^0$.
           The right panel includes the quanta with $\theta > 89.5^0$.}
  \label{peaks03}
\end{figure}

   Fig.~\ref{peaks02} demonstrates the details of the spectrum
presented in the top row in Fig.~\ref{peaks01}. Thus, the left
panel includes all the quanta emitted by a hot spot at $r =
10\,r_g$ with $\theta > 89^0$ at infinity (the mean value could be
counted as $85.5^0$ if the quanta distribution  would be uniform
there) but with much higher resolution than in Fig.~\ref{peaks01}.
The spectrum has four narrow distinctive maxima separated by lower
emission intervals. In the right panel, which includes all the
quanta with
 $\theta > 85.5^0$, the right complementary maxima is even higher
than the main one. The blue complementary maxima still remains
lower than its main counterpart, but it increases rapidly its
intensity with increasing inclination angle. It follows
immediately from the comparison of the left and right panels. The
"oscillation behavior" of the line profile between the maxima has
a pure statistical origin and is not caused by the physics
involved.

   As an illustration, a spectrum of an entire accretion disk at
high inclination angles in the Schwarzschild metric is shown in
Fig.~\ref{peaks03}. (In reality we have calculated geodesics for
the quanta trajectories in the Schwarzschild metric using the same
Eqs. (\ref{eq1})--(\ref{eq6}) as for a  Kerr metric, but assuming
there $a = 0.01$.) The emitting region (from 3 to $10\,r_g$) lies
as a whole in the innermost region of the $\alpha$-disk (the
detailed description of this model was given by
\cite{shasun,LipShak}).

   As  follows from  Fig.~\ref{peaks03}, the blue peak may consist of the two
components, whereas the red one remains unresolved.

\subsection{Discussion and conclusions}

    The complicated structure of the line profile at large inclination
angles is explained by the multiple images of some pieces of the
hot ring. We point out that the result was obtained in the
framework of GR without any extra physical and astrophysical
assumptions about the character of the radiation etc. For a Kerr
black hole we assume only that the radiating ring is circular and
narrow.

The problem of multiple images in the accretion disks and extra
peaks was first  considered by \citet{MPS93} (see also
\cite{Bao93,Bao92,BHO94,BHO96}). Using numerical simulations
\citet{MPS93} proved the statement for the Schwarzschild metric
and suggested that the phenomenon is applicable to a  Kerr metric
over the range of parameters that the authors have analyzed. They
noted also that it is necessary to perform detailed calculations
to confirm their hypothesis.  We verified and confirmed their
conjecture without any assumptions about a specific distribution
of surface emissivity or accretion disk model (see Fig.
\ref{peaks01},\ref{peaks02}).

 We confirmed also their conclusion  that extra peaks
are generated by photons which are emitted by the far side of the
disk, therefore we have a manifestation of gravitational lensing
in the strong gravitational field approach for GR.

Some  possibilities to observe considered features of spectral
line profiles were considered by \citet{MPS93, Bao93}. The authors
argued that there are non-negligible chances to observe such
phenomenon in some AGNs and X-ray binary systems. For example,
\citet{Bao93} suggested that NGC 6814 could be a candidate to
demonstrate such a phenomenon
(but \citet{Made93} found that the peculiar properties of NGC6814
are caused by a cataclysmic variable like an AM Herculis System).

However, it is clear that in general the probability to observe
objects where one could find such features of spectral lines is
small. Moreover, even if the inclination angle is very close to
$90^0$, the thickness of the disk ( shield or a torus around an
accretion disk) may not allow us to look at the inner part of
accretion disk.  Here, discuss astrophysical situations when the
the inclination angle of the accretion disk is high enough.

About 1\% of all AGN or microquasar systems could have an
inclination angle of the accretion disk  $> 89^0$. For example,
\citet{Kor96} found that NGC3115 has a very high inclination angle
about of $81^0$ (\citet{Kor92} discovered a massive dark object $M
\sim 10^9 M_\odot$(probably a massive black hole) in NGC3115).
Perhaps we have a much higher probability to observe such a
phenomenon in X-ray binary systems where black holes with stellar
masses could be. Taking into account the precession which is
actually observed for some X-ray binary systems (for example,
there is a significant precession of the accretion disk for the
SS433 binary system \cite{Cher02},\footnote{\citet{Shak72}
predicted that if the plane of an accretion disk is tilted
relative to the orbital plane of a binary system, the disk can
precess.} moreover since the inclination of the orbital plane is
high ($i\sim 79^0$ for this object) sometimes we may observe
almost edge-on accretion disks of such objects. Observations
indicated that there is  a strong evidence that the optically
bright accretion disk in SS433 is in a supercritical regime of
accretion. The first description of a supercritical accretion disk
was given by \citet{shasun}. Even now such a model is discussed to
explain observational data for SS433 \cite{Cher02}. We used the
temperature distribution from the Shakura -- Sunyaev model
\cite{shasun,LipShak} for the inner part of accretion disk to
simulate shapes of lines which could be emitted from this region
(Fig. \ref{peaks03}). Therefore, we should conclude that the
properties of spectral line shapes discovered by \citet{MPS93} are
confirmed also for such emissivity (temperature) distributions
which correspond to the Shakura -- Sunyaev model.

Thus, such properties of spectral line shapes are robust enough
with respect to wide variations of rotational parameters of black
holes and the surface emissivity  of accretion disks as it was
predicted by \citet{MPS93}. So, their conjecture  was confirmed
not only for the Kerr black hole case but also for other
dependences of surface emissivity of the accretion disk. A
detailed description of the analysis was given by
\citet{Zak_Rep03a}.

\section{MAGNETIC FIELDS IN AGNs AND MICROQUASARS}


     Magnetic fields play a key role in dynamics of accretion
discs and jet formation. \citet{Bis74,Bis76} considered a scenario
to generate superstrong magnetic fields near black holes.
According to their results magnetic fields near the marginally
stable orbit could be about $H \sim 10^{10} - 10^{11}$~G.
\citet{Kard95,Kard01} has shown that the strength of the magnetic
fields near supermassive black holes can reach the values $H_{max}
\approx 2.3\cdot 10^{10} M_9^{-1}$~G due to the virial
theorem\footnote{Recall that equipartition value of magnetic field
is $\sim 10^4$~G only.}, and considered a generation of
synchrotron radiation, acceleration of $e^{+/-}$ pairs and cosmic
rays in magnetospheres of supermassive black holes  at such high
fields. It is magnetic field, which plays a key role in these
models. Below, based on the analysis of iron $K_\alpha$ line
profile in the presence of a strong magnetic field, we describe
how to detect the field itself  or at least obtain an upper limit
of the magnetic field.

     General status of black holes is described in
a number of papers (see, e.g. \cite{Zak00} and references
therein). Since the matter motions indicate very high rotational
velocities, one can assume the $K_\alpha$ line emission arises in
the inner regions of accretion discs at distances $\sim (1\div
3)~r_g$ from the black holes. Let us recall that the innermost
stable circular for non-rotational black hole (which has the
Schwarzschild metric) is located at the distance $3\,r_g$ from the
black hole singularity. Therefore, a rotation of black hole could
be the most essential factor.

     Wide spectral lines are considered to be formed by
radiation  emitted in the vicinity of black holes.  If there are
strong magnetic fields near black holes these lines are split by
the field into several components. This phenomenon is discussed
below. Such lines have been found in microquasars, GRBs and other
similar objects \citep{Mira02a, Miller02}.

     To obtain an estimation of the magnetic field we simulate
the formation of the line profile for different values of magnetic
field. As a result we find the minimal $B$ value at which the
distortion of the line profile becomes significant. Here we use an
approach, which is based on numerical simulations of trajectories
of the photons emitted by a hot ring moving along a circular
geodesics near black hole, described earlier by
\cite{zakharov01,zakharov1,zak_rep1,zakharov5}.

\subsection{Influence of a magnetic field on the distortion
         of the iron $K_\alpha$ line profile}

    Here we consider the influence of a magnetic field on
the iron $K_\alpha$ line profile \footnote{We can also consider
X-ray lines of other elements emitted by the area of accretion
disc close to the marginally stable orbit; further we talk only
about iron $K_\alpha$ line for brevity.} and show how one can
determine the value of the magnetic field strength or at least an
upper limit.

    The profile of a monochromatic line
\citep{zak_rep1,zak_rep2} depends on the angular momentum of a
black hole, the position angle between the black hole axis and the
distant observer position, the value of the radial coordinate if
the emitting region represents an infinitesimal ring (or two
radial coordinates for outer and inner bounds of a wide disc). The
influence of accretion disc model on the profile of spectral line
was discussed by \citep{Zak_Rep03b}.

    We assume that the emitting region is located in
the area of a strong quasi-static magnetic field. This field
causes line splitting due to the standard Zeeman effect. There are
three characteristic frequencies of the split line that arise in
the emission. The energy of central component $E_0$ remains
unchanged, whereas two extra components are shifted by
$\pm \mu_B H$, where $\mu_B=\dfrac{e \hbar}{2m_{\rm
e}c}=9.273\cdot 10^{-21}$~erg/G is the Bohr magneton. Therefore,
in the presence of a magnetic field we have three energy levels:
$E_0-\mu_B H,~ E_0$ and $E_0+\mu_B H$. For the iron $K_\alpha$
line they are as follows: $E_0=6.4 - 0.58 \dfrac{H}{10^{11}\,{\rm
G}} $ keV, $E_0=6.4$ keV and $E_0=6.4 + 0.58
\dfrac{H}{10^{11}\,{\rm G}} $ keV.

   Let us discuss how the line profile changes when photons
are emitted in the co-moving frame with energy $E_0 (1+\epsilon)$,
but not with $E_0$. In that case the line profile can be obtained
from the original one by $1+\epsilon$ times stretching along the
x-axis which counts the energy. The component with $E_0
(1-\epsilon)$ energy should be $(1-\epsilon)$ times stretched,
respectively. The intensities of different Zeeman components are
approximately equal \citep{Fock76}. A composite line profile can
be found by summation the initial line with energy $E_0$ and two
other profiles, obtained by stretching this line along the
$x$-axis in $(1+\epsilon)$ and $(1-\epsilon)$ times
correspondingly. The line intensity depends on the direction of
the quantum escape with respect to the direction of the magnetic
field \citep{BLP89}. However, we neglect this weak dependence
(undoubtedly, the dependence can be counted and, as a result, some
details in the spectrum profile can be slightly changed, but the
qualitative picture, which we discuss, remains unchanged).

\begin{figure}
\vspace{-0.5cm}
\begin{minipage}[t]{6.3cm}
  \psfig{figure=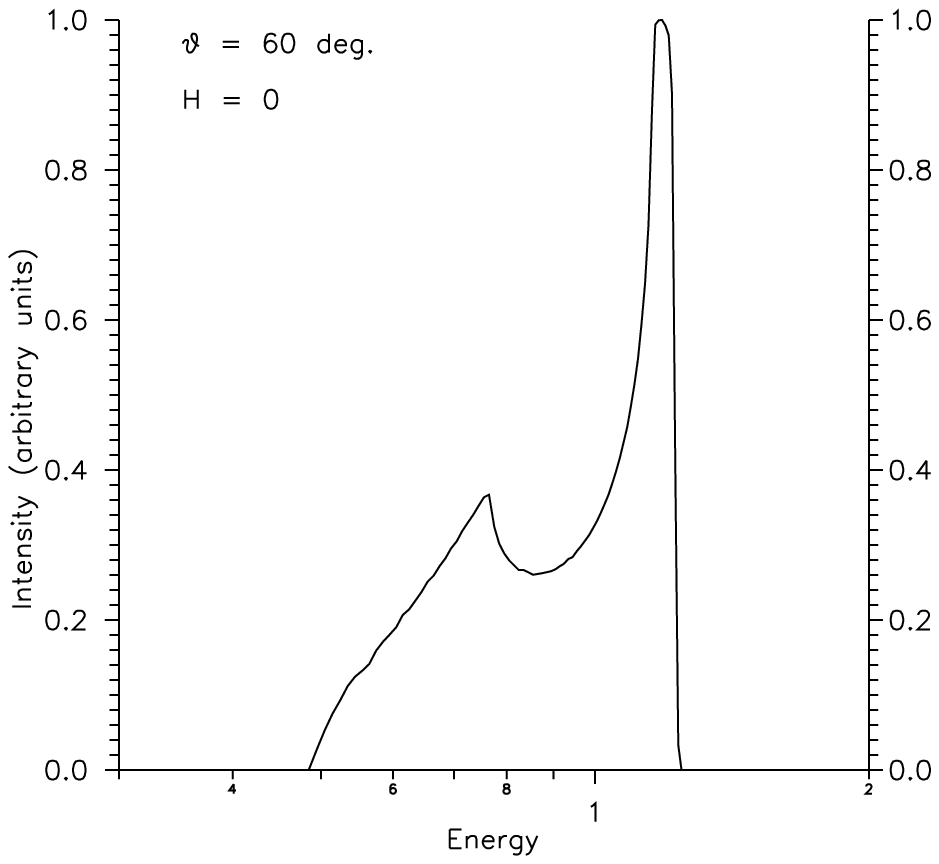,width=6.cm}
  \end{minipage}
\begin{minipage}[t]{6.3cm}
  \psfig{figure=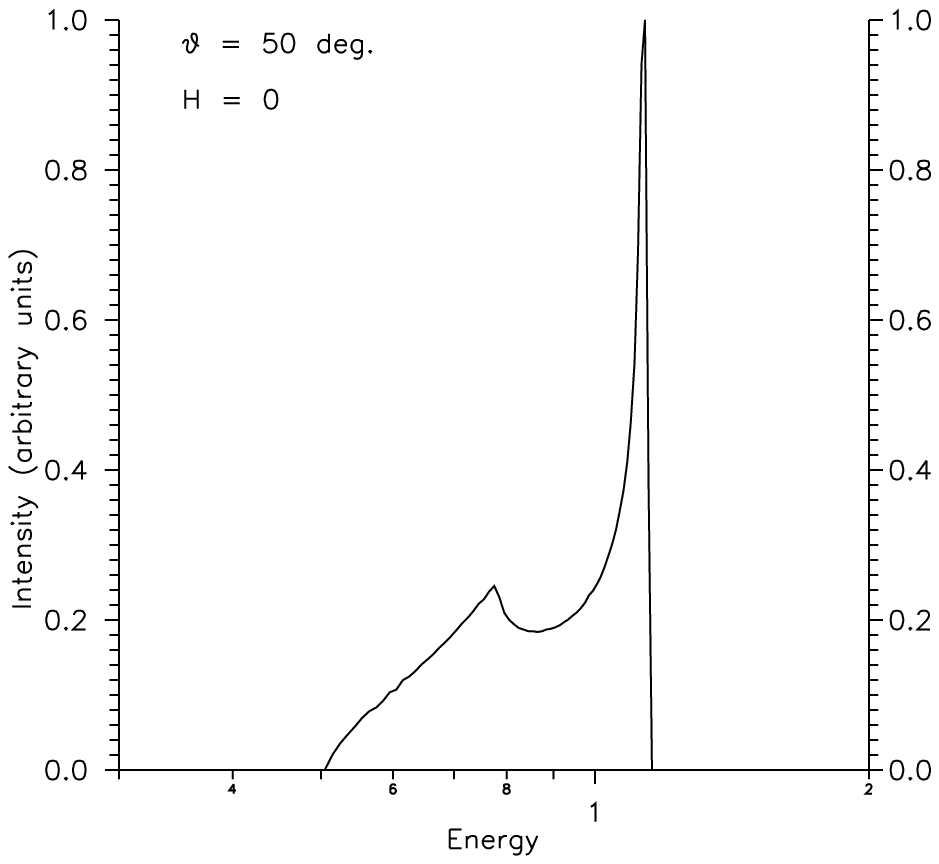,width=6.cm}
\end{minipage}
\vspace{-1cm}
  \caption{Profile of monochromatic spectral line, emitted
           by $\alpha$-disc in Schwarzschild metric for
           $r_{out} = 10\,r_g$, $r_{in} = 3\,r_g$ and
           inclination angles $\theta = 60^\circ$ (top panel)
           and $\theta = 50^\circ$ (bottom panel) with zero
           value of magnetic field. The line profile is represented as
           it is registered by a distant observer.}
  \label{zeeman01}
\end{figure}

\begin{figure}[t!]
\vspace{-.5cm}
  \psfig{figure=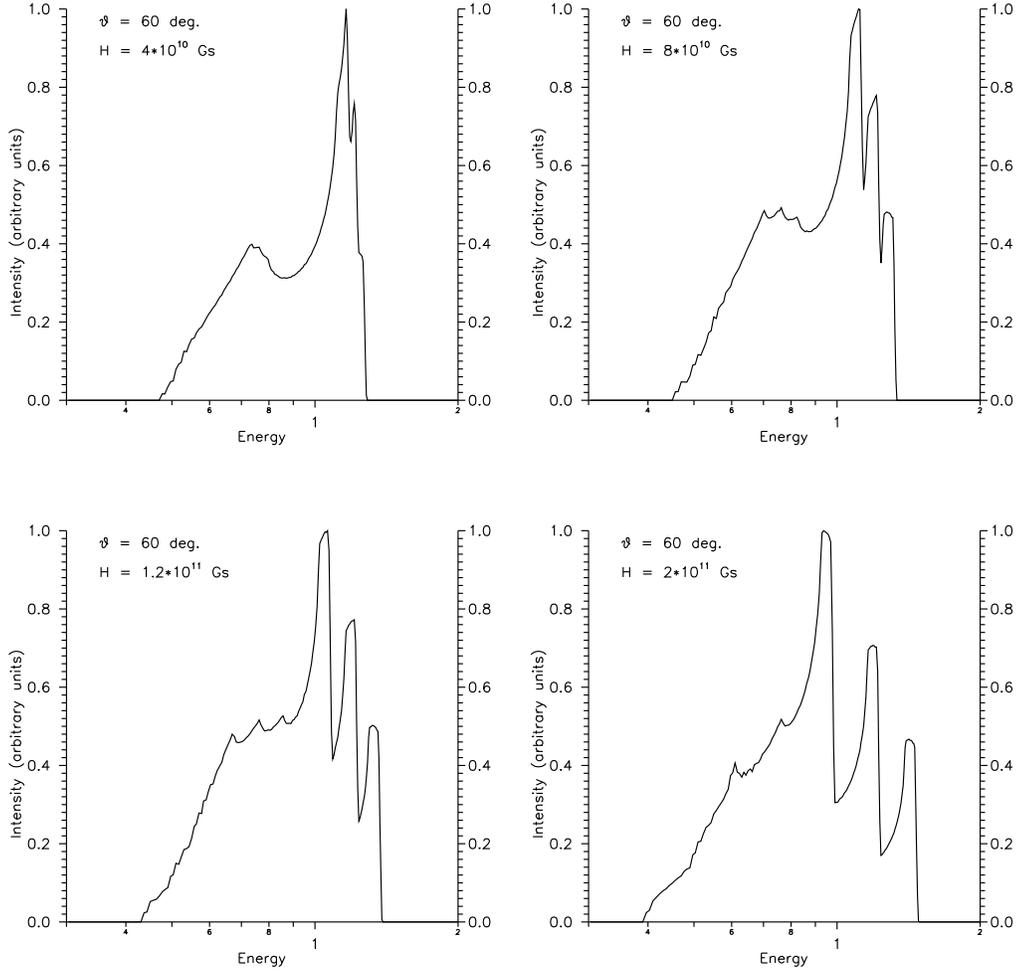,width=\textwidth}
\vspace{-1cm}
  \caption{Distortions of the line profile from the top panel
           in Fig.~\ref{zeeman01}, arising due to a
           quasi-static magnetic field existing in the disc.
           The Zeeman effect leads to the appearance of two
           extra components with the energies higher and lower
           than the basic one. The values of the magnetic
           field are shown at each panel.}
  \label{zeeman02}
\end{figure}

     Another indicator of the Zeeman effect is a significant
induction of the polarization of X-ray emission: the extra lines
possess a circular polarization (right and left, respectively,
when they are observed along the field direction) whereas a linear
polarization arises if the magnetic field is perpendicular to the
line of sight.
Despite of the fact that the measurements of polarization of X-ray
emission have not been carried out yet, such experiments can be
realized in the nearest future \citep{Cos01}.

\begin{figure}[t!]
\vspace{-0.5cm}
  \psfig{figure=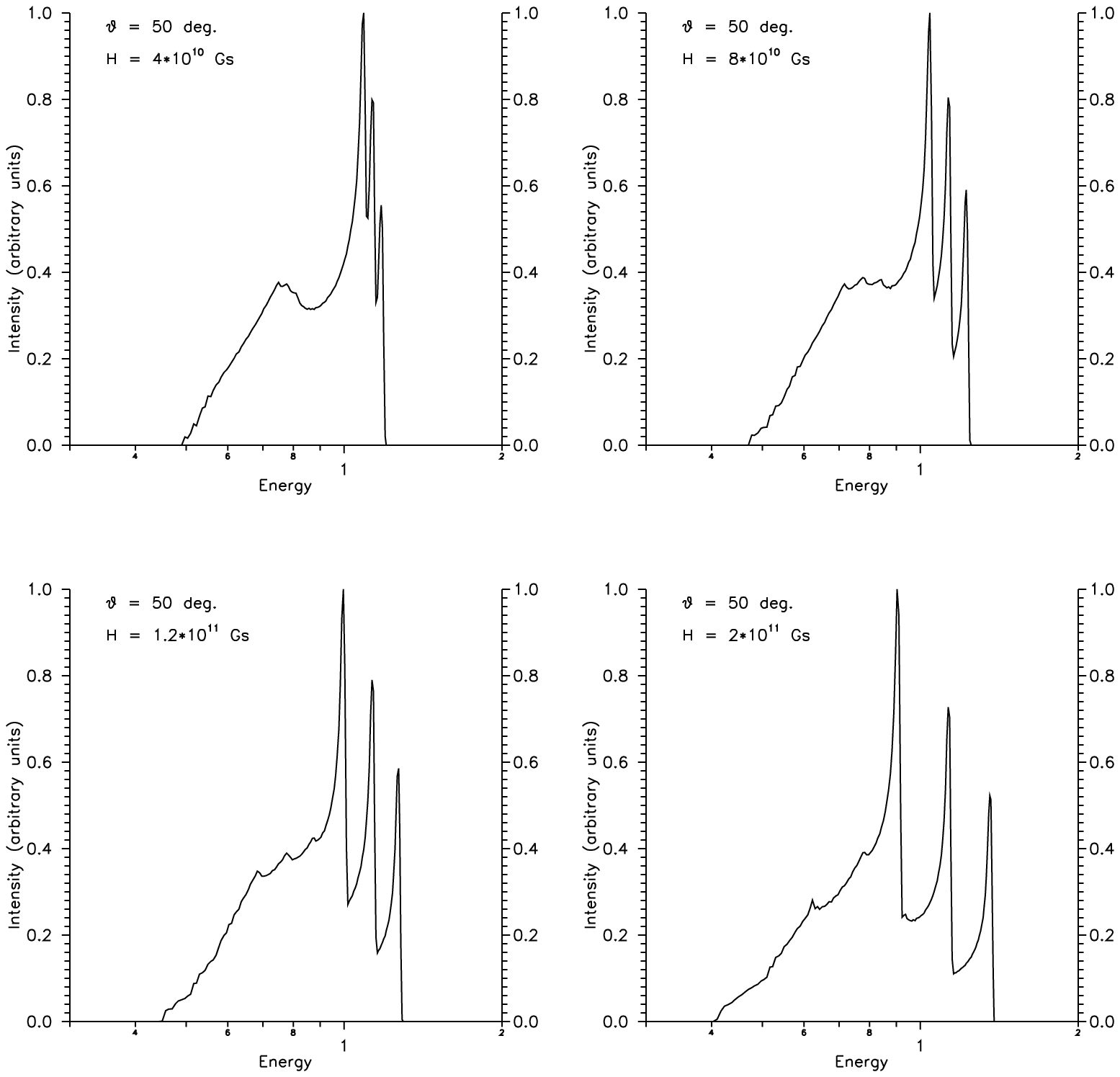,width=\textwidth}
\vspace{-1cm}
  \caption{The same as in Fig.~\ref{zeeman02},
           but for $\theta = 50^0$. The bottom panel on
           Fig.~\ref{zeeman01} demonstrates the same spectrum
           without magnetic field.}
  \label{zeeman03}
\end{figure}

\begin{figure}[t!]
\vspace{-0.5cm}
  \psfig{figure=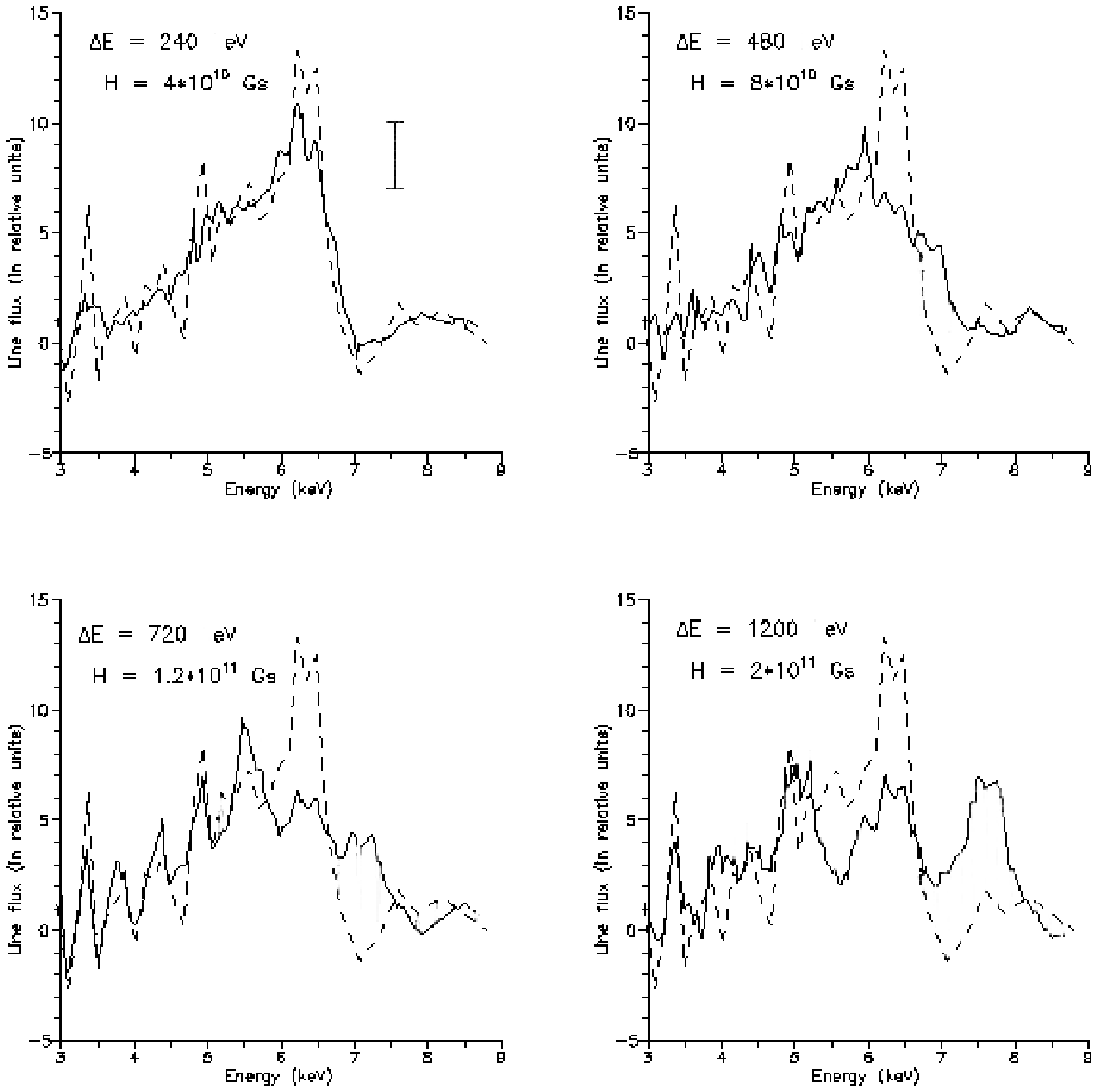,width=\textwidth}
\vspace{-0.5cm}
  \caption{Influence of a magnetic field on the observational
           data. The dashed line represents the ASCA observations
           of MCG-6-30-15 \citep{tanaka1}. The vertical bar
           in the top left panel corresponds to a typical error
           in observation data. Solid lines show possible
           profiles of $K_\alpha$ line in a presence of a magnetic
           field. The field value and the appropriate Zeeman
           splitting are indicated in each panel.}
  \label{zeeman05}
\end{figure}

    The line profile without any magnetic field is presented
in~Fig.~\ref{zeeman01} for different values of disc inclination
angles: $\theta = 60^0$ and $\theta = 50^0$ respectively. Note,
that at $\theta = 50^0$  the blue peak appears to be taller and
more narrow. Figs.~\ref{zeeman02},\ref{zeeman03} present the line
profiles for the same inclination angles and different values of
magnetic field: $H = 4\cdot 10^{10}$, $8\cdot 10^{10}$, $1.2\cdot
10^{11}$, $2\cdot 10^{11}$~G. At $H = 4\cdot 10^{10}$~G the shape
of spectral line does not practically differ from the one with
zero magnetic field. Three components of the blue peak are so thin
and narrow that they could  scarcely be distinguished
experimentally today. For $H < 4\cdot 10^{10}$~G and $\theta =
60^0$ the splitting of the line does not arise at all. At $\theta
= 50^0$ the splitting can still be revealed for $H = 3\cdot
10^{10}$~G, but below this value ($H < 3\cdot 10^{10}$~G) it also
disappears. With increasing the field the splitting becomes more
explicit, and at $H = 8\cdot 10^{10}$~G a faint hope appears to
register experimentally the complex internal structure of the blue
maximum.

     While further increasing the magnetic field
the peak profile structure becomes apparent and can be distinctly
revealed, however, the field $H = 2\cdot 10^{11}$~G is rather
strong, so that the classical linear expression for the Zeeman
splitting
\begin{equation}
      \Delta E = \pm \mu_B H
      \label{eq15a}
\end{equation}
should be modified. Nevertheless, we use Eq.(\ref{eq15a}) for any
value of the magnetic field, assuming that the qualitative picture
of peak splitting remains correct, whereas for $H = 2\cdot
10^{11}$~G the exact maximum positions may appear slightly
different. If the Zeeman energy splitting $\Delta E$ is of the
order of $E$, the line splitting due to magnetic fields is
described in a more complicated way. The discussion of this
phenomenon is not a point of this paper, our aim is to pay
attention to the qualitative features of this effect.

Let us discuss possible influence of high magnetic fields on real
observational data. We will try to estimate magnetic fields when
one could find the typical features of line splitting from the
analysis of the spectral line shape. Further we will choose some
values of magnetic field and simulate the spectral line shapes
from observational data for these values, assuming that these
observational data correspond to an object with no significant
magnetic fields. We will try to find signatures of the triple blue
peak analyzing the simulated data when magnetic fields are rather
high. Assuming that there are no essential magnetic fields
(compared to $10^{10}$~G) for some chosen object (for example, for
MCG 6-30-15) we could simulate the spectral line shapes for the
same objects but with essential
 magnetic fields.
     Fig.~\ref{zeeman05} demonstrates a possible influence
of the Zeeman effect on observational data. As an illustration we
consider the observations of iron $K_\alpha$ line which have been
carried out by ASCA for the galaxy MCG-6-30-15. They are presented
in Fig.~\ref{zeeman05} in the dashed curve. Let us assume that the
actual magnetic field in these data is negligible. Then we can
simulate the influence of the Zeeman effect on the structure of
observations and see if the simulated data (with a magnetic field)
can be distinguishable within the current accuracy of the
observations. The results of the simulated observation for the
different values of magnetic field are shown in
Fig.~\ref{zeeman05} in solid line. From these figures one can see
that classical Zeeman splitting in three components, which can be
revealed experimentally today, changes qualitatively the line
profiles only for rather high magnetic field. Something like this
structure can be detected, e.g. for $H = 1.2\cdot 10^{11}$~G, but
the reliable recognition of three peaks here is hardly possible.

    Apparently, it would be more correct to solve the
inverse problem:  try to determine the magnetic field in the disc,
assuming that the blue maximum is already split due to the Zeeman
effect. However, this problem includes too many additional
factors, which can affect on the interpretation. Thus, besides
magnetic field the line width depends on the accretion disc model
as well as on the structure of emitting regions. Problems of such
kind may become actual with much better accuracy of observational
data in comparison with their current state.

\subsection{Discussion}

    It is evident that duplication (triplication)
of a blue peak could be caused not only by the influence of a
magnetic field (the Zeeman effect), but by a number of other
factors. For example, the line profile can have two peaks when the
emitting region represents two narrow rings with different radial
coordinates (it is easy to conclude that two emitting rings with
finite widths separated by a gap, would yield a similar effect).
Despite of the fact that a multiple blue peak can be generated by
many causes (including the Zeeman effect as one of possible
explanation), the absence of the multiple peak can lead to a
conclusion about the upper limit of the magnetic field.

    It is known that neutron stars (pulsars) could have
huge magnetic fields. So, it means that the effect discussed above
could appear in binary neutron star systems. The quantitative
description of such systems, however, needs more detailed
computations.

    With further increasing of observational facilities it
may become possible to improve the above estimation. Thus, the
Constellation-X launch suggested in the coming decade seems to
increase the precision of X-ray spectroscopy as many as
approximately 100 times with respect to the present day
measurements \citep{weaver1}. Therefore, there is a possibility in
principle that the upper limit of the magnetic field can also be
100 times improved in the case when the emission of the X-ray line
arises in a sufficiently narrow region.

A detailed discussion of the magnetic field influence on spectral
line shapes for flat accretion flows was given by \cite{ZKLR02}
(see also paper \cite{Zakharov_03}) and for non-flat accretion
flows by \cite{ZMB03}. A possibility to observe mirages (shadows)
around black holes using space based interferometers (like
Radioastron space radio telescope) was discussed in paper
\cite{ZNDI_04}.


I would like to thank the organizers of SPIG-2004 Symposium for
their kind invitation to present the talk, profs. L. \v C.
Popovi\'c, M. S. Dimitrijevi\'c, F.~DePaolis, G.~Ingrosso and
J.~Wang  for very useful discussions and Dr. Z. Ma and S.V. Repin
for fruitful collaboration.



\end{document}